\documentclass[aps,twocolumn,showpacs,superscriptaddress]{revtex4}
\usepackage{graphicx,subfigure,amsmath,amssymb,multirow}

\begin{document}
\draft
\title{Transient currents in a molecular photo-diode}
\author{E. G. Petrov}
\affiliation{Bogolyubov Institute for Theoretical
Physics, National Academy of Sciences of Ukraine,
Metrologichna Street 14-B, UA-03680 Kiev, Ukraine}
\author{V. O. Leonov}
\affiliation{Bogolyubov Institute for Theoretical
Physics, National Academy of Sciences of Ukraine,
Metrologichna Street 14-B, UA-03680 Kiev, Ukraine}
\author{V. May}
\affiliation{Institut f\"{u}r Physik, Humboldt Universit\"{a}t zu Berlin, Newtonstrasse 15, D-12489 Berlin, Germany}
\author{P. H\"{a}nggi}
\affiliation{Institut f\"{u}r Physik, Universit\"{a}t Augsburg, Universit\"{a}tstrasse 1, D-86135
Augsburg, Germany}

\begin{abstract}
Light-induced charge transmission through a molecular junction (molecular diode) is studied
in the framework of a HOMO-LUMO model and in using a kinetic description. Expressions
are presented for the sequential (hopping) and direct (tunneling) transient current components
together with kinetic equations governing the time-dependent populations of the neutral and charged
molecular states which participate in the current formation. Resonant and off-resonant charge
transmission processes are analyzed in detail. It is demonstrated that the transient currents are
associated
with a molecular charging process which is initiated by photo excitation of the molecule.
If the coupling of the molecule to the electrodes is strongly asymmetric the transient currents can
significantly exceed the steady state current.
\end{abstract}


\maketitle

\section{Introduction}

The use of molecular nanostructures as diodes, transistors, switches, etc. is considered as one
possible way towards a further miniaturization of integrated circuits. Although
pioneering ideas in
this direction have been formulated more than 30 years ago \cite{car82,avi88} the detection
of current-voltage characteristics of a single molecule became only possible within the last 15
years (for an overview see \cite{met99,nit01,han02,joa05,cun05,gal07,che09,bur09}).
Up to date research still mainly focuses on an understanding of charge transmission in the
junction "electrode1-molecule-electrode2" (\textbf{1}-M-\textbf{2} system) where
a single molecule exhibits itself as an electron/hole transmitter.
It has been shown that at definite conditions the molecule is able to operate as a molecular diode.
For instance, during a coherent (elastic) electron tunneling in the biased \textbf{1}-M-\textbf{2}
system,  diode properties of the molecule appear only in the presence of a voltage drop across the
molecule. This conclusion is valid even at different contacts of the molecule with the electrodes
\cite{mrn02}. But, if an electron transmission is associated with incoherent electron transfer
processes   (inelastic tunneling or/and hopping), the molecular diode can originate from an unequal
coupling of the molecule to the electrodes. Just such a situation is considered in the present
paper. We show that a rectification effect can be observed even in the unbiased
\textbf{1}-M-\textbf{2} system where the  driving force of the electron transfer process is caused
by a photo excitation of the molecule.

Recent research addressed the use of organic molecules in molecular photo devices like
photo-diodes, photo-resistors, optical switches, and photo-amplifiers \cite{win07,kum08}.
For example, a light-controlled conductance switch based on a photochromic
molecule has been demonstrated \cite{mol09}. Moreover, single molecule luminescence
caused by the current through a molecular junction could be detected
\cite{qui03,ges04,don04,piv10}.

Theoretical estimates on the light-induced current and
current-induced light emission one can find in
\cite{Lehmann2002,Lehmann2003,LehmannJCP,galniz06,fai07,fai2011,buk08}. It has been shown that
a {\it dc}-current can be induced by an external {\it ac}-field either due to a
considerable difference between the electronic charge distributions within the molecular orbitals
(MOs), or if the amplitude of the electric field along one direction is larger than in the opposite
direction. The latter effect can be originated by a mixing of two laser pulses with
frequencies $\omega$ and $2\omega$ \cite{LehmannJCP,leh04,koh05,nat_nano,fra08}.
The generation of a {\it dc}-current
can be also achieved by an asymmetric distribution of molecular
energy levels caused by environmental fluctuations (such an asymmetry may induce a
ratchet current  \cite{kai08}). Besides the formation of a steady--state current
due to an optical excitation of the junction in the absence of an applied voltage,
a light-induced suppression of a current in the presence of an applied voltage
has been suggested as well \cite{kle06,kle07}. Note also the work on a
light-induced removal of the Franck-Condon blockade in a single-electron inelastic charge
transmission \cite{may08}.

While the examples mentioned above focus on steady--state properties of the junction,
also the formation of transient currents generated just after an alteration of the applied voltage
or by changing optical excitation attracted recent interest. The computations demonstrated that the
transient current in a molecular diode appearing just
after a sudden voltage switch-on or switch-off can significantly exceed
the steady--state value \cite{pk08,psmh11}. Such a behavior is caused by electron transfer
processes which are responsible for charging or discharge of the molecule and which are fast
compared to the processes that establish the steady state current.

It is the objective of the present work to study the  time-dependent behavior of the transient current
across a molecular junction in the absence of an applied voltage. In doing so we focus on the transient
current formed just after a fast switch--on of a \emph{cw}-optical excitation. Our analysis
allows to clarify the physical mechanisms which are responsible for the fast and the slow kinetic
phases of charge transmission through the molecular junctions.

The paper is organized as follows. General expressions for the sequential (hopping) and direct
(tunnel) current components in a molecular junction are given in the next section along
with the kinetic equations for the molecular state populations  and respective transfer
rates. In sec. III, a HOMO-LUMO description of the molecule is introduced to derive concrete
expressions for contact as well as inelastic tunnel rates. Expressions for the transient
photocurrent are presented in sec. IV. In sec. V, the results related to the off-resonant and
the resonant regime of current formation are discussed in details. Some concluding remarks
are presented in Sec. VI.

\section{Basic equations}

\subsection{Hamiltonian}

We introduce a model of the \textbf{1}-M-\textbf{2} molecular junction formed by two nonmagnetic
electrodes which are weakly coupled to the (nonmagnetic) molecule. The related
Hamiltonian of the system can be written as
%
\begin{equation}
H\,=\,H_e +H_m + H' + H_{f}(t)\; .
\label{hamlmr}
\end{equation}
The first term describes the Hamiltonian of
the ideal electrodes,
%
\begin{equation}
H_{e}\,=\,\sum_{{r\bf k}\sigma}\,E_{r \bf k}\,
 a^{+}_{r {\bf k}\sigma} a_{r {\bf k}\sigma}\,,
\label{hleads}
\end{equation}
where $E_{r \bf k}$ denotes the energy of a conduction band electron (with
wave vector $\bf k$) of the
$r(=1,2)$th electrode. For nonmagnetic electrodes and in the absence of
a magnetic field this energy
does not depend on the electron spin $\sigma$. Electron creation and
annihilation operators are denoted by
$a^{+}_{r {\bf k}\sigma}$ and  $a_{r {\bf k}\sigma}$, respectively. The expression
%
\begin{equation}
H_{m}=\sum_{M(N)}\,E_{M(N)}|M(N)\rangle\langle M(N)|\,
\label{hammol1}
\end{equation}
defines the Hamiltonian of the molecule, where $E_{M(N)}$ denotes the energy
of  the molecule in state $|M(N)\rangle$. The quantum
number $M$ labels the actual electronic, vibrational, and spin state of
the molecule; $N$ denotes the number of electrons in the molecule.
The third term in eq. (\ref{hamlmr}) reads
%
\begin{displaymath}
H'=
\sum_{r{\bf k}\sigma}\,\sum_{N,MM'}\,[V_{M'(N+1);r{\bf k}\sigma M(N)}\,
\end{displaymath}
\begin{equation}
\times |M'(N+1)\rangle\langle M(N)|\,a_{r{\bf k}\sigma} +h.c.]\, .
\label{int1}
\end{equation}
It describes the molecule--electrode interaction with the
matrix element $V_{M'(N+1);r{\bf k}\sigma M(N)}=
\langle M'(N+1)|V_{tr}|r{\bf k}\sigma M(N)\rangle$ characterizing the
electron exchange ($V_{tr}$ is the electron transfer operator).
The interaction of the molecule with an external \emph{cw}-field is
written in the standard form
%
\begin{equation}
H_{f}(t)=-{\bf E}(t)\sum_{MM'N}\,\textbf{d}_{M'M}|M'(N)\rangle\langle M(N)|\,
\label{field}
\end{equation}
where $\textbf{E}(t)$ is the electric component of the periodic field and
$\textbf{d}_{M'(N)M(N)}$ is the transition dipole matrix element between
different states of the molecule.

\subsection{Sequential and direct components of an electron current}

The current across the electrode $r$ is given by
%
\begin{equation}
I_r(t)=e\,(\delta_{r,1}-\delta_{r,2})\,\dot{N}_r(t)\,
\label{curdef}
\end{equation}
where $e=-|e|$ is the electron charge,  $\dot{N}_r(t)=
\sum_{{\bf k}\sigma}\,\dot{P}(r{\bf k}\sigma;t)$
denotes the electron flow from the $r$th electrode,  and $P(r{\bf k}\sigma;t)$
is the population of
the single-electron band state. For stationary charge
transmission the number of electrons leaving one of the electrodes is
identical with the number
of electrons arriving at the other electrode,
i.e. $\dot{N}_1(t)=-\dot{N}_2(t)=\text{const}$. In the
nonstationary regime, however, $\dot{N}_1(t)$ and $\dot{N}_2(t)$ may be
quite different from each
other so that $I_1(t)\neq I_2(t)$ (cf. refs. \cite{pk08,psmh11}).

Nonequilibrium density matrix (NDM) theory \cite{akh81,blu96,may04} is quite
suitable to achieve a unified description of elastic (coherent) as well as
inelastic (hopping and
incoherent) charge transmission in the molecular junctions. Such description
allows one to express the transfer rates characterizing the noted
transmission via the set of molecule-electrode
couplings and transmission gaps. In refs.
\cite{psmh11,pet06,pmh06,maku08,pet11}, the NDM theory has been used to derive
kinetic equations for the single electron populations
$P(r{\bf k}\sigma;t)$ and the molecular populations $P(M(N);t)$. Just these
equations determine the evolution of the current components in
the molecular junctions.

In the presence of an external \emph{cw}-field, the
derivation procedure becomes more complicated. If, however, the interaction, eq.
(\ref{field}), acts as a perturbation only,
the calculation of electron transfer rates associated with the interaction, eq.
(\ref{int1}), can be carried out by ignoring the molecule-field interaction.
The condition that permits one to consider the interaction
(\ref{field}) as a perturbation, reduces to the inequality
%
\begin{equation}
\omega^2\gg |{\bf E}\textbf{d}_{M'M}|^2/\hbar^2
\label{fnon}
\end{equation}
where $|{\bf E}|$ is the amplitude of the \emph{cw}-field. Owing to
the condition (\ref{fnon}), only  single photon transitions with
frequency $\omega = (1/\hbar)|E_{M(N)}-E_{M'(N)}|$ will
support charge transfer processes in the \textbf{1}-M-\textbf{2} device.
The derivation of kinetic equations for the populations
$P(r{\bf k}\sigma;t)$ and $P(M(N);t)$ remains identical with that already presented in
\cite{psmh11,pet06,pmh06,maku08,pet11}. Therefore, we do not repeat the derivation here.
We only mention that for the considered weak
molecule-electrode coupling a unified description
of charge transmission is achieved by using the transition operator
$\hat{T}=H'+H'\hat{G}(E)H'$ (note that the
electrodes stay in equilibrium). The matrix elements
$\langle a|\hat{T}|b\rangle$
determine the transitions between the states $b$ and $a$ on the energy shell
$E=E_a=E_b$ \cite{dav76}, where the $E_a$ and $E_b$ are energies referring to the
Hamiltonian $H_0=H_e+H_m$. The
Green's  operator $\hat{G}(E)=(H_0+H'+i0^+)^{-1}$ is defined by the
Hamiltonian of the whole \textbf{1}-M-\textbf{2} system in the absence of
molecule-field interaction.
The first term of $\hat{T}$ is responsible for a single electron hopping
between the molecule and the attached electrodes while the second term
results in a direct one-step electron transition between the electrodes.
Besides, the operator
$H'\hat{G}(E)H'$ is responsible for a specific electron--pair transition between the
molecule and the electrodes. Respective transfer rates are presented
in refs. \cite{psmh11,pet11}. The mechanism of
electron--pair transitions has been applied earlier to explain the nonlinear
electron transport through a single-level
quantum dot \cite{lei09}. For such a transport the repulsion between
the transferred electrons in the dot is compensated by the
voltage bias. In the present paper, a light-induced
charge transmission is considered in an unbiased molecular junction.
Therefore, the energies of twofold charged molecular states are arranged high enough
to only give a negligible contribution
to the current. It means that the pair--electron transfer
processes become unimportant and, thus,
our study is limited to single electron transmission
processes. As a result, the current through the $r$th electrode has
two components
%
\begin{equation}
I_r(t)=I^{(r)}_{seq}(t) +I_{dir}(t)\,.
\label{curdet1}
\end{equation}
The sequential component,
%
\begin{displaymath}
I^{(r)}_{seq}(t)=|e|\,(-1)^{r+1}\sum_{N,MM'}
(\chi^{(r)}_{M(N)\rightarrow M'(N+1)}
\end{displaymath}
\begin{equation}
-\chi^{(r)}_{M(N)\rightarrow M'(N-1)})
P(M(N);t)\,
\label{seq1}
\end{equation}
is defined by single electron jumps through the contact region between
the electrode surface and the molecule. Respective hopping transfer rates
can be referred to the contact forward (electrode-molecule) and contact
backward (molecule-electrode) rates which read
%
\begin{displaymath}
\chi^{(r)}_{M(N)\rightarrow M'(N+1)}
= \frac{2\pi}{\hbar}\sum_{{\bf k}\sigma}\,
|V_{M'(N+1);r{\bf k}\sigma  M(N)}|^2\,
\end{displaymath}
\begin{equation}
\times \,f_r(E_{r{\bf k}})\,\delta[E_{r{\bf k}}+E_{M(N)}-E_{M'(N+1)}]
\label{cf1}
\end{equation}
and
%
\begin{displaymath}
\chi^{(r)}_{M(N)\rightarrow M'(N-1)}
= \frac{2\pi}{\hbar}\sum_{{\bf k}\sigma}\,
|V_{M'(N-1)r{\bf k}\sigma;  M(N)}|^2\,
\end{displaymath}
\begin{equation}
\times \,[1-f_r(E_{r{\bf k}})]\,\delta[E_{r{\bf k}}+
E_{M'(N-1)}-E_{M(N)}]\,.
\label{cb1}
\end{equation}
[In eqs. (\ref{cf1}) and (\ref{cb1}), $f_r(E_{r{\bf k}})=
\{\,\exp{[(E_{r{\bf k}} -\mu_{r})/k_BT]} +1\}^{-1}$
is the Fermi distribution function with $\mu_{r}$ being the chemical
potential for the $r$th electrode.] Contact forward and backward
rates are responsible, respectively,  for reduction and oxidation
of the molecule by the $r$th electrode. In mesoscopic physics, a similar
type of hopping processes is classified as an electron tunneling between the lead
and the dot \cite{gla05}.

The current component
%
\begin{equation}
I_{dir}(t)=|e|\,\sum_{N,MM'}S^{(dir)}_{MM'}(N)\,P(M(N);t)
\label{dir1}
\end{equation}
is formed by an interelectrode electron transfer at
which the molecule mediates a charge transmission without alteration of
its charge. Such process is defined by the electron flows
%
\begin{equation}
S^{(dir)}_{MM'}(N)=Q_{1M(N)\rightarrow 2M'(N)}-Q_{2M(N)\rightarrow 1M'(N)}\,
\label{dirflow}
\end{equation}
where the transfer rates
%
\begin{displaymath}
Q_{r M(N)\rightarrow r'M'(N)}
\end{displaymath}
\begin{displaymath}
=\frac{2\pi}{\hbar}
\sum_{{\bf k}\sigma,}\,\sum_{{\bf k}'\sigma'}\,f_r(E_{r{\bf k}})\,
[1-f_{r'}(E_{r'{\bf k}'})]\,
\end{displaymath}
\begin{displaymath}
\times |\langle M'(N)r'{\bf k}'\sigma'|H'\hat{G}(E)H'|r{\bf k}
\sigma M(N)\rangle|^2
\end{displaymath}
\begin{equation}
\times
\delta[E_{r{\bf k}}+E_{M(N)} -E_{r'{\bf k}'}-E_{M'(N)}]\,
\label{mmrate}
\end{equation}
characterize a distant electron transmission from the ${\bf k}\sigma$
band states of the $r$th electrode to the ${\bf k}'\sigma'$ band states of
the $r'$th electrode. Such transmission appears as a direct single-step
elastic (at $M'(N)=M(N)$) or inelastic (at $M'(N)\neq M(N)$) interelectrode
electron tunneling. Since the operator $H'$
is responsible for transitions accompanied by an alteration of molecular
charge, the mediation of the tunneling
transmission occurs via the formation of intermediate molecular states
$\tilde{M}(N+1)$ and $\tilde{M}(N-1)$ which differ from the initial, $M(N)$
and final, $M'(N)$ charge states. In the contrast to the sequential
(hopping) transmission where similar states are really populated, the
noted intermediate states are not populated and only acts as virtual states. In mesoscopic physics,
such type of transmission
refers to co-tunneling \cite{gla05}. In the respective terminology the
direct current component, eq. (\ref{dirflow}), results
as a contribution of partial currents associated with different
co-tunneling channels. The realization of a particular
channel is controlled by the probability $P(M(N);t)$ to find a molecule in
the $M(N)$th stay. [Examples of electron transmission along the channel
pathways that include the empty and occupied MOs can be found
in \cite{pmh05,pet07}.] Thus, the
direct tunneling can be referred to as co-tunneling which is controlled by
kinetic charging and recharge of the molecule (via electron jumps through
the contact region). This circumstance has been already noted in \cite{pzmh06}.
In the presence of the \emph{cw}-field, an additional control
occurs through the population of the excited molecular state.

\subsection{Kinetic equations for the molecular populations}

It follows from eqs. (\ref{seq1}) and (\ref{dir1}) that each charge
transmission route (sequential or direct) includes electron transfer
channels related to the molecular states $M(N)$. The contribution of
the $M(N)$th channel to the route is weighted by the
molecular population $P(M(N);t)$, which satisfies the normalization condition
%
\begin{equation}
\sum_{NM}\,P(M(N);t)=1\,.
\label{norm1}
\end{equation}
Following the derivation procedure presented in refs.
\cite{psmh11,pet11,lux10} and bearing in mind the fact that the interactions
(\ref{int1}) and (\ref{field}) are considered as perturbations,  we can see
that evolution of the $P(M(N);t)$ is determined by the balance like kinetic
equation
%
\begin{displaymath}
\dot{P}(M(N);t)=-\sum_{M'N'}[{\mathcal K}_{M(N)\rightarrow M'(N')}\,P(M(N);t)
\end{displaymath}
\begin{equation}
-{\mathcal K}_{M'(N')\rightarrow M(N)}\,P(M'(N');t)] \; .
\label{nm2}
\end{equation}
The transfer rate
%
\begin{displaymath}
{\mathcal K}_{M(N)\rightarrow M'(N')}= \sum_{r}\,
(\delta_{N',N+1}+\delta_{N',N-1})\,
\end{displaymath}
\begin{displaymath}
\times\chi^{(r)}_{M(N)\rightarrow M'(N')}+\delta_{N,N'}\,[K^{(f)}_{M(N)\rightarrow M'(N)}
\end{displaymath}
\begin{equation}
+\sum_r(1-\delta_{r,r'})\,
Q_{r M(N)\rightarrow  r'M'(N')}]\,
\label{nm23}
\end{equation}
specifies the  transition from the state $M(N)$ to the state $M'(N')$ in
the molecule. Such transition is caused by the molecule-electrode interaction
(\ref{int1}) through the contact and distant transfer rates
(eqs. (\ref{cf1}, (\ref{cb1}), (\ref{mmrate})) as well as by the molecule-field
interaction (\ref{field}). Respective rates of optical excitation and
de-excitation are
%
\begin{displaymath}
K_{M(N)M'(N)}^{(f)}=\frac{2\pi}{\hbar}\,
|{\bf E}\textbf{d}_{M'(N)M(N)}|^2\,
\end{displaymath}
\begin{equation}
\times [L_{M'(N)M(N)}(\omega)+L_{M'(N)M(N)}(-\omega)]\,.
\label{kf1}
\end{equation}
We introduced
$L_{M'(N)M(N)}(\omega)=(1/2\pi)\,(\kappa_{M(N)}+
\kappa_{M'(N)})\{[\hbar\omega -( E_{M'(N)}-
E_{M(N)})]^2 +(\kappa_{M(N)}+\kappa_{M'(N)})^2/4\}^{-1}$, where
$\kappa_{M(N)}/2$ denotes the molecular level broadening caused by
electron-phonon interaction as well as interaction of the molecule with the electrodes (for more
details see \cite{lux10}).

\section{Charge transfer processes in the HOMO-LUMO model}

Next, the  hopping rates, eqs. (\ref{cf1})
and (\ref{cb1}) as well as the distant transfer rate, eq. (\ref{mmrate}),
all determining the net electron flow through the junction, are further
specified along the rate, eq. (\ref{kf1}) characterizing
the efficiency of excitation and de-excitation of the molecule.
We use a model of the \textbf{1}-M-\textbf{2} system where only the
highest occupied and the lowest unoccupied molecular orbitals (HOMO ($H$)
and LUMO ($L$), respectively) are considered. The HOMO-LUMO model is
\begin{figure}
\includegraphics[width=7.3cm]{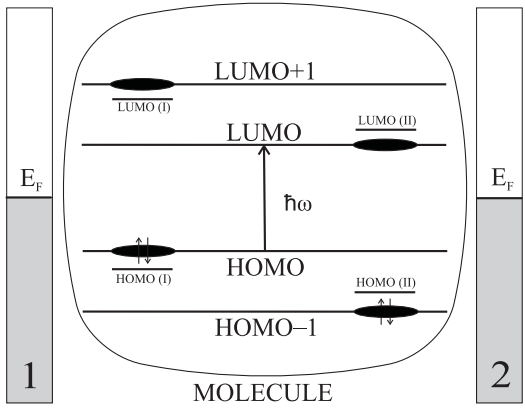}
\caption{Possible position of the frontier MOs in the molecule with two terminal sites I and II.
Intersite coupling transforms the site MOs into extended HOMO, HOMO-1 and LUMO, LUMO+1. Spots
indicate  the main location of electron density within the extended MOs.
}
\label{fig01}
\end{figure}
suitable to study charge transmission in the molecular
junctions. As an example, note the pioneer work of Aviram and
Ratner \cite{avi74} where the mechanism of current formation includes a
participation of HOMO and LUMO levels belonging to the
donor and acceptor sites of the molecule. Recently, a similar model
(with chromophoric donor and acceptor sites) has been used for the
description of transient dynamics in a molecular junctions \cite{vol11}.
In this model, a transient electronic current is formed due to an
optical excitation associated with the HOMO-LUMO transition in the
donor site.

In the present paper, we use a model where extended HOMOs (LUMOs) are
formed from HOMOs (LUMOs) belonging to the terminal molecular sites I and II
coupled to one another by interior bridging groups.
Let the HOMO($n$) and the LUMO($n$) refer to the terminal site $n$(=I,II).
Following from the coupling between the
sites,  the extended HOMO and HOMO -1 (LUMO and LUMO +1)
represent a mixture of the HOMO(I) and HOMO(II) (LUMO(I) and LUMO(II)).
If the intersite coupling does not strongly modify the electron distribution
across the molecule,  the maxima of electron density in the
HOMO, HOMO -1, LUMO, and LUMO +1
correspond to  electron densities located in the vicinity of the respective
sites, cf. Fig. \ref{fig01}. Therefore, the coupling
of the HOMO to electrode 1 is assumed to be much stronger than the
similar coupling to electrode 2. The opposite case is valid for
the coupling of the LUMO to the same electrodes. This configuration as
represented in Fig. \ref{fig01} can be realized if, for instance, the
HOMO(I)/LUMO(I) and HOMO(II)/LUMO(II) refer to the $\pi$-electrons of
aromatic groups coupled to each other by the bridging
$\sigma$-bonds (to avoid a noticeable mixture between the
$\pi$-electrons belonging the sites I and II).
If the energy  $\hbar\omega$ of the external \emph{cw}-field coincides with the
energy of the optical HOMO-LUMO transition, then the formation of the
photocurrent can be mainly associated  with two  frontier MOs (HOMO
and LUMO). In this case, the rates of optical excitation and de-excitation
are determined by  eq. (\ref{field}).

For the subsequent analysis we assume that the Coulomb interaction between
excess electrons (or holes) occupying the molecule in the course of charge
transfer, is so large that the molecule
can only stay in its neutral ground (or excited) state, in its oxidized state
and in its reduced state. These states are denoted as $M_0=M(N_G)$,
$M_*=M'(N_G)$, $M_+=M(N_G-1)$ and $M_-=M(N_G+1)$. Here, $N_G$ is the number of
electrons if the molecule is in its neutral state.
If the maxima of electron location at the
HOMO and LUMO  are in the vicinity of the spaced sites I and II,
cf. Fig. \ref{fig01}, one can suppose that the exchange interaction between the
unpaired electrons occupying the HOMO and the LUMO becomes small. This
allows one to ignore the exchange splitting between the singlet, $M_*(S)$ and triplet,
$M_*(Tm), (m=0,\pm 1)$ states of the excited molecule. Accordingly, the electron spin projections
can be taken as good quantum numbers.
Therefore, the four-fold degenerated excited state $M_*$ can be characterized
either by molecular spin states ($M_*=M_*(S),M_*(Tm)$) or by spin
projections $\sigma_H$ and $\sigma_L$ of unpaired electrons occupying the
frontier MOs ($M_*=M_*(\sigma_H,\sigma_L)$). At a negligible exchange
interaction, both sets of spin quantum numbers lead to identical
results. Moreover, the states $M_+=M_+(\sigma_H)$ and $M_-=M_-(\sigma_L)$
are twofold degenerated.

To specify the energies $E_{M(N)}$ entering the molecular Hamiltonian
(\ref{hammol1}) and the matrix elements in the interaction expression
(\ref{int1}) we introduce the following notation of the Hamiltonian
%
\begin{displaymath}
H_{m}\,=\,\sum_{j}\,\sum_{\sigma}\,(\epsilon_j
+U_j\,c^+_{j-\sigma}c_{j-\sigma}
\end{displaymath}
\begin{equation}
+\frac{1}{2}\,\sum_{j'(\neq j)}\,\sum_{\sigma'}\,U_{jj'}\,
c^+_{j'\sigma'}c_{j'\sigma'})\,c^+_{j\sigma}c_{j\sigma}\,
\label{hmol}
\end{equation}
and of the electron transfer coupling (cf. \cite{pzmh06,mwl91,mwl93,hsw02})
%
\begin{equation}
V_{tr}\,=\,\sum_{j}\,\sum_{r{\bf k}\sigma}\,
(\beta_{jr {\bf k}}\,c^+_{j\sigma}a_{r {\bf k}\sigma}
+\beta^*_{jr {\bf k}}\, a^{+}_{r {\bf k}\sigma}c_{j\sigma})\,.
\label{hld-m}
\end{equation}
In eq. (\ref{hmol}) the $\epsilon_j$ are the energies of an electron
occupying the $j(=H,L)$th MO.
The strength of the Coulomb interaction between two electrons is defined by $U_j$
if both electrons occupy the $j$th MO. If the electrons belong to different MOs Coulomb
interaction is measured by $U_{jj'}$. The operators $c^{+}_{j\sigma}$ and
$c_{j\sigma}$ create or annihilate an electron in the molecule, and $\beta_{jr \bf k}$
characterizes the coupling of the $j$th MO to the $r \bf k$th band state of the electrode.

According to the Hamiltonian, eq. (\ref{hmol}), the molecular energies
$E_{M(N)}=E_{\alpha}$, ($\alpha=0,*,+,-$)  follow as:
%
\begin{displaymath}
E_0=2\epsilon_H +U_H\,,\;\;E_*=\epsilon_H +\epsilon_L+U_{LH}\, ,
\end{displaymath}
\begin{equation}
E_-=2\epsilon_H +\epsilon_L+ U_H+2U_{LH}\,, \;\;E_+=\epsilon_H\,.
\label{molene}
\end{equation}
Here, $\epsilon_H $  and $\epsilon_L $  are the energies of an electron occupying the frontier MOs
while $U_H$ and $U_{HL}$ are the Coulomb parameters.
Concerning the matrix elements entering eq. (\ref{int1}), all of them are expressed by
the couplings $\beta_{Hr \bf k}$ or $\beta_{Lr \bf k}$. We have, for example, $\langle
M_0|V_{tr}|M_+(\sigma_H)r{\bf k}\sigma\rangle=\beta_{Hr \bf k}\delta_{-\sigma,\sigma_H}$ and
$\langle M_0r{\bf k}\sigma|V_{tr}|M_-(\sigma_L)\rangle=\beta^*_{Lr \bf k}\delta_{\sigma,\sigma_L}$.

\subsection{Contact rate constants}

\begin{figure}
\includegraphics[width=7.3cm]{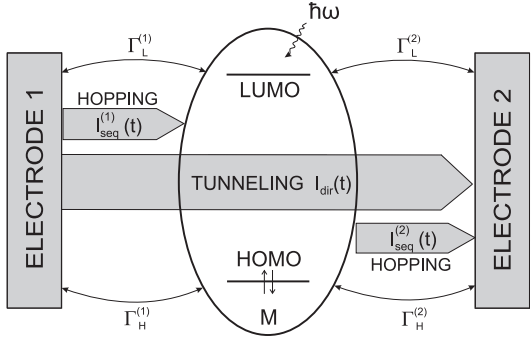}
\caption{HOMO-LUMO scheme related to the electron transfer through the
\textbf{1}-M-\textbf{2} molecular junction.
The width parameters $\Gamma_j^{(r)}$ characterize the efficiency of contact
electron jumps as well as of the direct (tunneling) electron transfer. The
sequential current components $I_{seq}^{(1)}(t)$  and $I_{seq}^{(2)}(t)$ can
differ from each other if the junction is transient regime.
}
\label{fig02}
\end{figure}

\begin{figure}
\includegraphics[width=7.3cm]{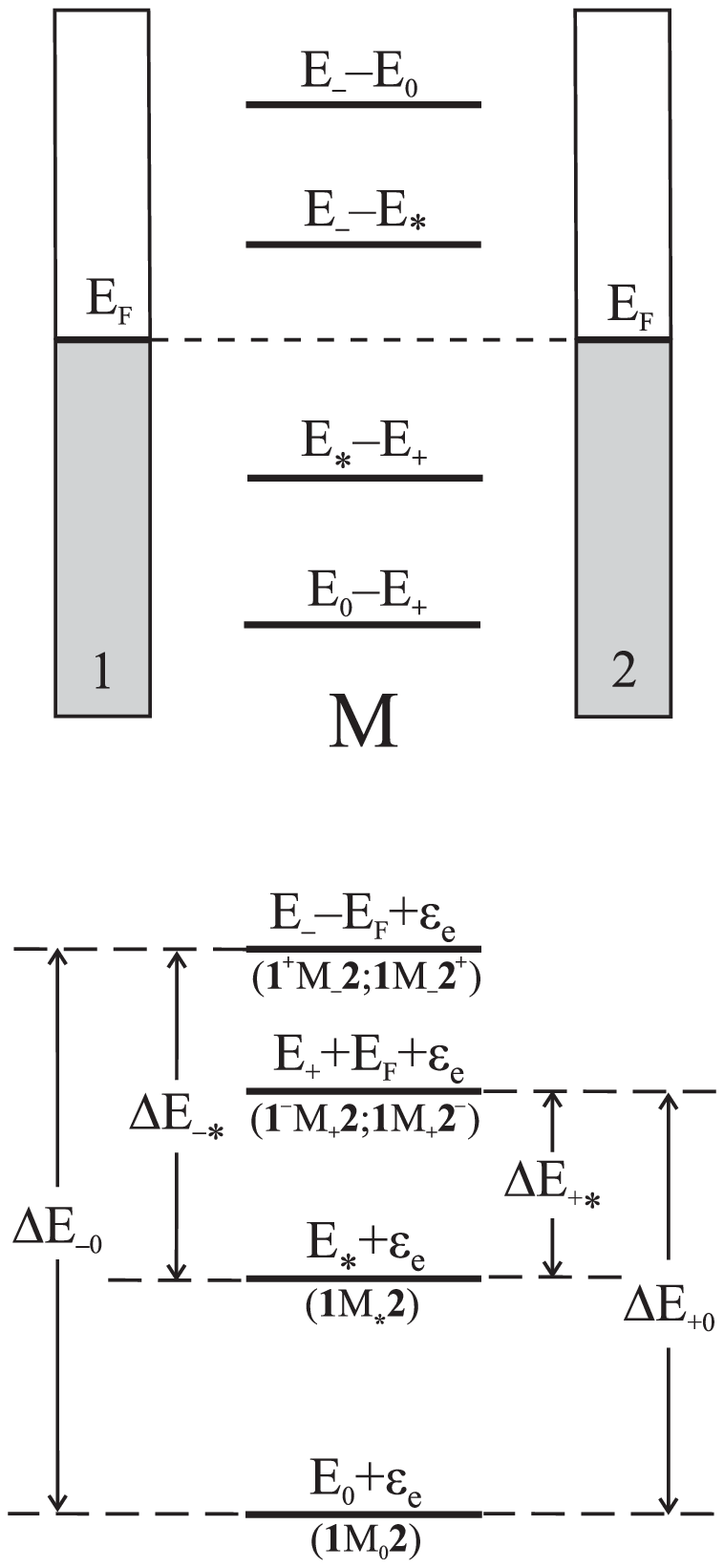}
\caption{Charging energies $E_0-E_+$, $E_*-E_+$, $E_- E_0$ and $E_--E_*$
(upper panel) and transmission gaps related to the charged molecular states $M_+$ and $M_-$
(lower panel).
}
\label{fig03}
\end{figure}

Noting the structure of the transition matrix elements, the so-called wide
band approximation \cite{foot1} enables one to express the hopping
transfer rates (\ref{cf1}) and (\ref{cb1}) by  contact rate constants
$K^{(r)}_{\alpha\alpha'}$. For instance, we get
$\chi^{(r)}_{M_*(\sigma_H,\sigma'_L)\rightarrow
M_-(\sigma_L)}=\delta_{\sigma'_L,\sigma_L}\,K^{(r)}_{*-}$. According to
the used HOMO-LUMO model the forward contact rate constants takes the form
%
\begin{displaymath}
K^{(r)}_{0\,-}
\simeq (1/\hbar)\,\Gamma^{(r)}_{L}\,
N(\Delta E_{-0})\,,\;\;
\end{displaymath}
\begin{displaymath}
K^{(r)}_{*\,-}
\simeq (1/\hbar)\,\Gamma^{(r)}_{H}\,
N(\Delta E_{-*})\,,\;\;
\end{displaymath}
\begin{displaymath}
K^{(r)}_{+\,0}
\simeq (1/\hbar)\,\Gamma^{(r)}_{H}\,
N(\Delta E_{0+})\,,\;\;
\end{displaymath}
\begin{equation}
K^{(r)}_{+\,*}
\simeq (1/\hbar)\,\Gamma^{(r)}_{L}\,
N(\Delta E_{*+})\, .
\label{contrate}
\end{equation}
The quantities
%
\begin{equation}
\Gamma^{(r)}_{j}
\simeq 2\pi\,\sum_{{\bf k}}\,|\beta_{j{r\bf k}}|^2\,
\delta (E-E_{{r\bf k}})\,
\label{width1}
\end{equation}
characterize electron hopping between the $j$th MO and the $r$th electrode
(cf. Fig. \ref{fig02}), whereas the distribution function
%
\begin{equation}
N(\Delta E_{\alpha'\alpha})=
[\exp{(\Delta E^{(r)}_{\alpha'\alpha}/k_BT)}
+1]^{-1}\, .
\label{disfer1}
\end{equation}
determines the influence of temperature on the hopping processes via the transmission
gaps
%
\begin{equation}
\Delta E_{+0(*)}=(E_++E_F)-E_{0(*)}\,
\label{gap+}
\end{equation}
and
%
\begin{equation}
\Delta E_{-0(*)}=E_--(E_{0(*)} + E_F)\,.
\label{gap-}
\end{equation}
Backward contact rate constants which characterize
the transition of an electron from the molecule to the $r$th electrode are
connected with the forward ones, eq. (\ref{contrate}), by the relation
%
\begin{equation}
K^{(r)}_{\alpha\,\alpha'}=
K^{(r)}_{\alpha'\,\alpha}\,\exp{(-\Delta E_{\alpha'\alpha}}/k_BT)\,.
\label{contrateb}
\end{equation}
The physical meaning of the transmission gaps can be easily deduced from their definition.
Since $E_--E_{0}$  and $E_+-E_0$ are the electron charging and electron
discharging energies (with
respect to the molecule being in its ground neutral state), respectively, the inequalities
$E_--E_{0}>E_F$ and $E_0-E_+< E_F$ have to be fulfilled in the
unbiased \textbf{1}-M-\textbf{2} system with identical electrodes
(cf. the upper panel in Fig. \ref{fig03} where $\mu_1=\mu_2=E_F$).
Therefore, the  gaps $\Delta E_{-0}$ and $\Delta E_{+0}$  are both
positive. When the molecule is in the excited state,
then respective  charging and discharging energies,
$E_--E_{*}$ and $E_+-E_*$, can be higher or lower than the Fermi level and, thus,
transmission gaps $\Delta E_{-*}$ and $\Delta E_{+*}$ can become positive or
negative. [One possible case with $E_--E_{*}>E_F$ and $E_*-E_+< E_F$, is
presented at the upper panel in Fig. \ref{fig03}.]

An additional interpretation of the transmission gaps follows from a
comparison of electron energies belonging the whole \textbf{1}-M-\textbf{2}
system \cite{psmh11,pet06}.
Let ${\mathcal E}_e$ be the energy of electrons in the electrodes. In the case
of a neutral molecule  the energy of the whole system is
$E(\textbf{1}M_{0(*)}\textbf{2})=E_{0(*)}+\mathcal{E}_{e}$. During a charge
transmission process the number of electrons in the
system is conserved. Therefore, the energies of the system with the
oxidized and reduced molecule are, respectively,
$E(\textbf{1}^-M_+\textbf{2})=E(\textbf{1}M_+\textbf{2}^-)=
E_++ \mathcal{E}_e +E_F$
and $E(\textbf{1}^+M_-\textbf{2})=E(\textbf{1}M_-\textbf{2}^+)=
E_-+ \mathcal{E}_e -E_F$.
Therefore, the gaps (\ref{gap+}) and (\ref{gap-}) correspond to
the difference between the above noted energies,
i.e.  $\Delta E_{+0(*)} = E(\textbf{1}^-M_+\textbf{2})-
E(\textbf{1}M_{0(*)}\textbf{2})
= E(\textbf{1}M_+\textbf{2}^-)-E(\textbf{1}M_{0(*)}\textbf{2})$ and
$\Delta E_{-0(*)} = E(\textbf{1}^+M_-\textbf{2})-
E(\textbf{1}M_{0(*)}\textbf{2})
= E(\textbf{1}M_-\textbf{2}^+)-E(\textbf{1}M_{0(*)}\textbf{2})$
(cf. Fig. \ref{fig03} lower panel). Such interpretation of the
transmission gaps is quite suitable for the analysis of the transmission
processes in the molecular junctions.

The sign of the transmission gap defines the electron transfer along a given
transmission channel. For instance, if
$\Delta E_{+*}$ is positive, then the transition $M_{+}\rightarrow M_{*}$  caused by an electron
injected into the molecule, requires a thermal activation i.e. it proceeds in an
off-resonant regime. If $\Delta E_{+*}<0$, however, the
$M_{+}\rightarrow M_{*}$ transition does not require any thermal activation and, thus, becomes
practically independent on the absolute value of $\Delta E_{+*}$. Consequently, the
electron hopping takes place in a resonant regime.

\subsection{Inelastic tunnel rate constant}

Eq.  (\ref{mmrate}) for the direct (tunnel) transfer rate indicates that in
the absence of an applied voltage any elastic electron tunneling between
identical electrodes disappears. We study in the following an inelastic
tunneling event which is accompanied by the intramolecular transition
$M_*\rightarrow M_0$. To derive a respective rate expression we first
consider the effect of the molecule-electrode  coupling. Since it is not too strong, its presence
may be accounted for by a shift $\Delta E_{M(N)}$ of
the molecular energies as well as by a level broadening
%
\begin{displaymath}
\Gamma_{M(N)}/2=\pi\sum_r\,\sum_{M'}\sum_{{\bf k}\sigma}\,\big\{
|V_{M(N)r{\bf k}\sigma;M'(N+1)}|^2
\end{displaymath}
\begin{displaymath}
\times \delta[E_{M'(N+1)}-E_{M(N)}-E_{r{\bf k}}]
\end{displaymath}
\begin{equation}
+|V_{M(N);M'(N-1)r{\bf k}\sigma}|^2\delta[E_{M(N)}-E_{M'(N-1)}
-E_{r{\bf k}}]\big\}\,.
\label{wp}
\end{equation}

Accordingly, the Green's operator, determining the general
transition amplitude is defined by these shifted and broadened molecular
energies. Due to the weak molecule-electrode coupling one can omit
the energy shift $\Delta E_{M(N)}$.
Taking the Hamiltonian, eq. (\ref{hmol}), then, in the
framework of the HOMO-LUMO model, the unperturbed molecular energies,
eq. (\ref{molene}) are expressed via the single-electron energies
$\epsilon_{H}$  and $\epsilon_{L}$ as well as via
the Coulomb parameters $U_H$ and $U_{LH}$. The broadenings,  eq. (\ref{wp}) are defined
by the single-electron level broadenings
$\Gamma _{j}/2$. The quantities
%
\begin{equation}
\Gamma _{j}=\sum_{r}\,\Gamma^{(r)}_{j}\,,\;\;\;\;(j=H,L)\,,
\label{broad}
\end{equation}
are obtained as the sum of the width parameters $\Gamma^{(r)}_{j}$, eq. (\ref{width1}). Thus, the
described formulation
of the Green's operator $\hat{G}(E)$ and the introduction of the width
parameters results in the following rate expression
%
\begin{displaymath}
Q_{r M_*(\sigma_L,\sigma_H)\rightarrow r' M_0}=Q^{(rr')}_{*\,0}
\end{displaymath}
\begin{displaymath}
\simeq\frac{1}{\pi\hbar}\,\Big[\frac{\Gamma^{(r')}_{L}
\Gamma^{(r)}_{H}}{\Gamma_{+}}\,\big(\varphi_{+\rightarrow 0}-
\varphi_{+\rightarrow *}\big)
\end{displaymath}
\begin{equation}
+\frac{\Gamma^{(r')}_{L}
\Gamma^{(r)}_{H}}{\Gamma_{-}}\,\big(\varphi_{-\rightarrow 0}-
\varphi_{-\rightarrow *}\big)\Big]\,.
\label{ineltun}
\end{equation}
Here we used  $\Gamma_{+}=\sum_r\,(\Gamma^{(r)}_{H}+
2\Gamma^{(r)}_{L})$ and $\Gamma_{-}=\sum_r\,(\Gamma^{(r)}_{L}+
2\Gamma^{(r)}_{H})$. According to eq. (\ref{ineltun}) the  regime of
inelastic  tunneling transfer is governed by the quantities
%
\begin{equation}
\varphi_{\alpha'\rightarrow\alpha}={\rm arc\,tn}\,
(2\Delta E_{\alpha'\alpha}/\Gamma_{\alpha'})\,.
\label{phi}
\end{equation}
For the weak molecule-electrode coupling under consideration the width
parameters do not exceed $10^{-3}$ eV. Accordingly, we have
$|\Delta E_{\alpha'\alpha}/\Gamma_{\alpha'}|\gg 1$, and one can use the
asymptotic form
$\varphi_{\alpha'\rightarrow\alpha}\approx (\pi/2)({\rm sign}\Delta
E_{\alpha'\alpha})-(\Gamma_{\alpha'}/2\Delta E_{\alpha'\alpha})$.
It follows the particular relation
$\varphi_{+\rightarrow 0}-\varphi_{+\rightarrow *}
\approx (\pi/2)(1-{\rm sign}\Delta E
_{+*}) +[(\Gamma_+/2\Delta E _{+ *})-(\Gamma_+/2\Delta E_{+0})]
\approx (\pi/2)[(1-{\rm sign}\Delta E
_{+*}) +(\Gamma_+/\pi \Delta E _{+*})]$. Taking $\Delta E _{+ *}>0$ then
$\varphi_{+\rightarrow 0}-\varphi_{+\rightarrow *}
\approx (\Gamma_+/\pi \Delta E_{+*})\ll 1$. In the contrary case
$\Delta E _{+*}<0$ we find $\varphi_{+\rightarrow 0}-\varphi_{+\rightarrow *}
\approx \pi$ what is much larger then
the respective expression deduced for $\Delta E _{+*}>0$. Note also that at
the resonant regime of tunnel electron transmission the
$\varphi_{+\rightarrow 0}-\varphi_{+\rightarrow *}$ is independent of the
actual value of the transmission  gap.

\subsection{Rate equations for integral molecular populations}

\begin{figure}
\includegraphics[width=7.3cm]{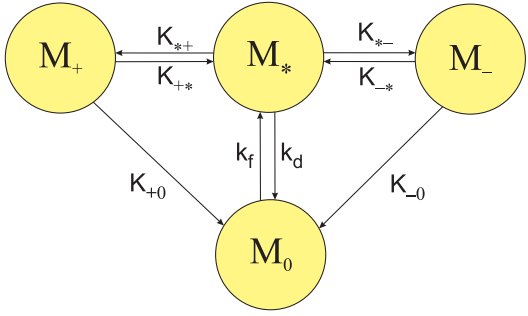}
\caption{(color online) Kinetic scheme of the transfer processes
occurring in the molecular junction in the absence of an
applied voltage (for further details see text).
}
\label{fig04}
\end{figure}

The kinetics in the junction are dominated by
sequential processes which are characterized by contact rate constants
$K^{(r)}_{\alpha\alpha'}$. But, the direct inelastic  electron tunneling
responsible for the transitions between the excited
and the ground states of the neutral molecule, is also important. For
instance, the distant rate constants $Q^{(rr')}_{*0}$ describe the
nonradiative decay of the excited molecule.
The kinetic scheme depicted in Fig. \ref{fig04} illustrates the possible
transition processes in the junction including the \emph{cw}-optical
excitation of the molecule. All rates indicated in Fig. \ref{fig04}
characterize the transitions with
the participation of the degenerated molecular states $M_*$, $M_+$ and $M_-$.
Therefore, it is convenient to introduce integral molecular populations
%
\begin{displaymath}
P(*;t)=\sum_{\sigma_H,\sigma_L}\,P(M_*(\sigma_H,\sigma_L);t)\,,
\end{displaymath}
\begin{displaymath}
P(+;t)= \sum_{\sigma_H}\,P(M_+(\sigma_H);t)\,,\;\;
\end{displaymath}
\begin{equation}
P(-;t)=\sum_{\sigma_L}\,P(M_-(\sigma_L);t)\,,.
\label{prob2}
\end{equation}
[Note also that  $P(*;t)=P_*(S;t)+\sum_{m=0,\pm 1}\,P_*(Tm);t)$.]
The quantities introduced in eq. (\ref{prob2}) along with the population $P(0;t)\equiv P(M_0;t)$
obey the normalization condition
%
\begin{equation}
\sum_{\alpha=0,*,+,-}\,P(\alpha;t)=1\,
\label{norm2}
\end{equation}
which corresponds to eq. (\ref{norm1}). Based on the introduction of integral
populations, the general kinetic equations (\ref{nm2}) reduce to the following set
of rate equations
%
\begin{displaymath}
\dot{P}(0;t)=-k_fP(0;t)
+K_{+0}P(+;t)
\end{displaymath}
\begin{displaymath}
+K_{-0}P(-;t)+ k_dP(*;t)\,,
\end{displaymath}
\begin{displaymath}
\dot{P}(+;t)=-(K_{+0}+2K_{+*})P(+;t)
+K_{*+}P(*;t)\,,
\end{displaymath}
\begin{displaymath}
\dot{P}(-;t)=-(K_{-0}+2K_{-*})P(-;t)
+K_{*-}P_*(t)\,,
\end{displaymath}
\begin{displaymath}
\dot{P}(*;t)=-(K_{*+}+K_{*-}+k_d)P(*;t)
\end{displaymath}
\begin{equation}
+2K_{+*}P(+;t)+2K_{-*}P(-;t)+k_fP_0(t)\,.
\label{intkin1}
\end{equation}
Here, we introduced the recharge transfer rates
%
\begin{equation}
K_{\alpha\alpha'}=K^{(1)}_{\alpha\alpha'} + K^{(2)}_{\alpha\alpha'}
\label{rechrate}
\end{equation}
which are expressed by the sum of contact rate constants.
The rate constant $k_f=K^{(f)}_{0*}$, eq. (\ref{kf1}) characterizes the optical
transition between molecular singlet states $M(N)=M_0$ and $M'(N)=M_*(S)$.
Accordingly, the overall decay rate from the four-fold degenerated
excited state follows as
%
\begin{equation}
k_{d}=k_f/4+Q_{*0} \; .
\label{drate}
\end{equation}
We defined
%
\begin{equation}
Q_{*0}=Q^{(12)}_{*0}+Q^{(21)}_{*0}\,
\label{tr+}
\end{equation}
as the component caused by the coupling of the molecule to the electrodes.
The concrete expression for $Q_{*0}$ can be deduced from eq. (\ref{ineltun}).
It reads
%
\begin{equation}
Q_{*0}\simeq \frac{1}{2\hbar}\,\big(\Gamma^{(1)}_{H}
\Gamma^{(2)}_{L}+\Gamma^{(2)}_{H}
\Gamma^{(1)}_{L}\big)\,R
\label{dirtr+}
\end{equation}
with
%
\begin{displaymath}
R=\Gamma^{-1}_{+}\,[(1-{\rm sign}\Delta E_{+*})+(\Gamma_+/\pi\Delta E_{+*})]
\end{displaymath}
\begin{equation}
+\Gamma^{-1}_{-}\,[(1-{\rm sign}\Delta E_{-*})+(\Gamma_-/\pi\Delta E_{-*})]\,.
\label{ineltun3}
\end{equation}

\section{The Photocurrent}

Already in the absence of an applied voltage a photocurrent,
eq. (\ref{curdet1}) has to be expected. According to the used HOMO-LUMO
model and by noting the  eqs. (\ref{cf1}), (\ref{cb1}), (\ref{contrate}) and
(\ref{contrateb}) for the sequential current component, eq. (\ref{seq1}),
we find
%
\begin{displaymath}
I^{(r)}_{seq}(t)=(\delta_{r,1}-\delta_{r,2})\,I_0\,\pi\hbar\,
\{[K^{(r)}_{*-}\,P(*;t)
\end{displaymath}
\begin{displaymath}
+(K^{(r)}_{+0}+2K^{(r)}_{+*})\,P(+;t)]-[K^{(r)}_{*+}\,P(*;t)
\end{displaymath}
\begin{equation}
+(K^{(r)}_{-0}+2K^{(r)}_{-*})\,P(-;t)]\}\, ,
\label{cseq2}
\end{equation}
where $I_0=|e|/\pi\hbar\times 1{\rm eV}\approx 77.6\,\mu$A is the current
unit \cite{am}. It follows from eq. (\ref{cseq2}) that
the charge transmission along the sequential route is determined by
hopping (contact) rate constants (\ref{contrate}) and (\ref{contrateb}).
The time-dependent behavior of this current component
is determined by the molecular state populations  $P(+;t)$, $P(-;t)$ and
$P(*;t)$. The direct component of the current is formed by
the inelastic tunnel electron transmission along the channel related to
the excited molecule. The expression for the direct component follows
from eqs. (\ref{dir1}), (\ref{dirflow}), and (\ref{ineltun}) and takes
the form
%
\begin{equation}
I_{dir}(t)=I_0\,\pi\hbar\,
 S_{*0}\,P(*;t)\,.
\label{cdir2}
\end{equation}
Here, we introduced
%
\begin{displaymath}
S_{*0}=Q^{(12)}_{*0}-Q^{(21)}_{*0}
\end{displaymath}
\begin{equation}
\simeq \frac{1}{2\hbar}\,\big(\Gamma^{(1)}_{H}
\Gamma^{(2)}_{L}-\Gamma^{(2)}_{H}
\Gamma^{(1)}_{L}\big)\,R
\label{ineltun2}
\end{equation}
what represents the net tunnel electron flow ($R$ has been introduced in
eq. (\ref{ineltun3}) ).
Eqs. (\ref{cdir2}) and (\ref{ineltun2}) show that, the sign of the direct
current component (in the HOMO-LUMO model) is determined by the sign of
$\Gamma^{(1)}_{H}\Gamma^{(2)}_{L}-\Gamma^{(2)}_{H}\Gamma^{(1)}_{L}$.
Moreover, the time-dependent behavior of the direct current component is
determined by the population  $P(*;t)$.

\begin{figure}
\includegraphics[width=7.3cm]{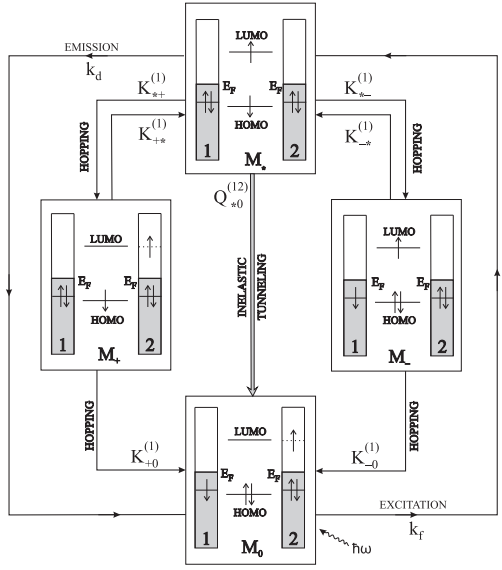}
\caption{Transfer routes of the light-induced  interelectrode
$\textbf{1}\textbf{2}\rightarrow\textbf{1}^+\textbf{2}^{-}$ electron transmission
including the participation of the charged molecular states $M_+$ and $M_-$. Within the transmission
along the sequential route, the charged states are populated while the same states
participate in a virtual form only if the transfer proceeds along the tunnel route.
}
\label{fig05}
\end{figure}

The scheme of the electron transfer routes, as displayed in
Fig. \ref{fig05}, offers the opportunity
to analyze further details of the current formation.
Charge transitions are represented by $\textbf{1}\textbf{2}\rightarrow
\textbf{1}^{+}\textbf{2}^-$. According to Fig. \ref{fig05} a possible
current formation results from the decay of the excited molecular state.
This is possible via the sequential route (including the formation of the
charged molecular states $M_+$ and $M_-$) as well as via the
direct route $M_*\rightarrow M_0$. Both routes include two (left and right)
transmission channels.
The left sequential channel, $\textbf{1}M_*\textbf{2}\rightleftarrows
\textbf{1}M_+\textbf{2}^-\rightarrow \textbf{1}^+M_0\textbf{2}^-$,  proceeds
across the charged molecular state $M_+$. This state is formed by an electron
hopping from the LUMO to the electrode 2 (with contact rate constant
$K^{(2)}_{*+}$) after which another electron hops from the electrode 1 to
the HOMO (with contact rate constant $K^{(1)}_{+0}$). In summary, we have
the transition $\textbf{1}\textbf{2}\rightarrow \textbf{1}^+\textbf{2}^-$.
This transition is also achieved by an electron transmission along the
right sequential channel  $\textbf{1}M_*\textbf{2}\rightleftarrows
\textbf{1}^+M_-\textbf{2}\rightarrow \textbf{1}^+M_0\textbf{2}^-$ which
includes the charged molecular state $M_-$. The $M_-$ state is formed by
an electron hopping from electrode 1 to the HOMO. The transfer of an electron
from the LUMO to electrode 2 returns the molecule to its neutral ground state
$M_0$. The respective sequential charge transfer steps are
characterized by the contact hopping rates $K^{(1)}_{*-}$ and $K^{(2)}_{-0}$.
It is important to underline that during the electron transmission along the sequential route, the
current formation is accompanied by a molecular recharging, i.e by a population of  the intermediate
charged  molecular states $M_+$ and $M_-$.

The second type of transfer route shown Fig. \ref{fig05}, refers to the
direct (distant) interelectrode electron transfer
$\textbf{1}\textbf{2}\rightarrow \textbf{1}^+\textbf{2}^-$ which is
accompanied by the $M_*\rightarrow M_0$ transition in the molecule.
In contrast to the sequential route, electron transmission along the direct
route $\textbf{1}M_*\textbf{2}\rightarrow \textbf{1}^+M_0\textbf{2}^-$ does
not result in a change of the charged molecular states populations (these
states only participate as virtual states). Thus, a direct electron
transfer constitutes an inelastic tunneling event of an electron between the
electrodes. The related distant transfer rate $Q_{*0}^{(12)}$ is defined in
eq. (\ref{ineltun}). Since a reverse route
$\textbf{1}M_0\textbf{2}\leftarrow \textbf{1}^-M_*\textbf{2}^+$ is formed in
a similar way and is characterized by the rate $Q_{*0}^{(21)}$, the direct
current component is proportional to the net electron flow $S_{*0}$, eq.
(\ref{ineltun2}).

\section{Results and discussion}

The eqs. (\ref{curdet1}), (\ref{cseq2}), and (\ref{cdir2}) allow one to describe
the time-dependent evolution of the
photocurrent in the molecular junction starting with the switch-on of a
\emph{cw}-optical excitation (at $t=0$) and extending up to the formation of
a steady-state current $I_{st}=I_1(t\gg\tau_{st})=I_2(t\gg\tau_{st})$  where
$\tau_{st}$ is the characteristic time of the steady state formation. The
kinetic schemes drawn in the Figs. \ref{fig04} and \ref{fig05} show that a
control of the light-induced electron transfer is achieved via the transitions
between electronic states $M_0, M_*, M_+$ and $M_-$ of
the molecule. Here, the charged molecular states $M_+$ and
$M_-$ are of particular importance since those participate in the transmission
channels formation related to the sequential and the direct electron transfer
routes. As far as the charged states population is determined by the relation
between forward and backward contact rate constants, eqs. (\ref{contrate})
and (\ref{contrateb}), the position of the energy levels in the
\textbf{1}-M-\textbf{2} system predetermines the specific form of  the
interelectrode
$\textbf{1}\textbf{2}\rightarrow \textbf{1}^+\textbf{2}^-$ and
$\textbf{1}^-\textbf{2}^+\leftarrow \textbf{1}\textbf{2}$
electron transmission.

\begin{figure}
\includegraphics[width=7.3cm]{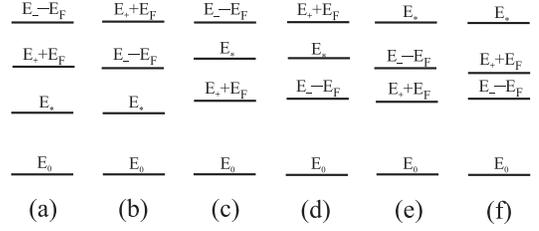}
\caption{Feasible  energy levels of the molecular junctions in the absence of an applied voltage.
$E_0,E_*,E_+$ and $E_-$ are the molecular energies and $E_F$ denotes the electrode
Fermi energy.
}
\label{fig06}
\end{figure}

Possible arrangements of the molecular junctions energy levels are depicted
in Fig. \ref{fig06}. For the sake of simplicity, the electron energy
$\mathcal{E}_e$ of the electrodes is omitted (cf. Fig. \ref{fig03}, lower panel where this energy is
presented).
If the junction energy with the neutral molecule in its excited state is
below the energy valid if the molecule is in a charged state, then only an
off-resonant regime of light-induced electron transmission becomes possible
(cases (a) and (b)). The cases (c) and (d) correspond to charge transmission
with a single resonant channel (either the oxidized or the
reduced molecule is involved). Two further transmission channels
are realized if  the junction energy with the neutral molecule in its
excited state exceeds the energies realized for the oxidized or the reduced
molecule (cases (e) and (f)).

The further analysis will be based on the eqs. (\ref{curdet1}), (\ref{cseq2}),
(\ref{cdir2}) for the current as well as the eqs. (\ref{contrate}) and
(\ref{contrateb}) defining the contact rate constants.
Additionally, the expressions (\ref{tr+}) - (\ref{ineltun3}) for the tunnel
decay rate $Q_{*0}$ and eq. (\ref{ineltun2}) for net electron flow $S_{*0}$
are taken into account. The time-dependent evolution of the integral molecular
populations $P(\alpha;t)$ are determined by the rate equations (\ref{intkin1})
where the recharge rate constants are given by  eq. (\ref{rechrate}). Initial
conditions for the populations are found from a solution of the rate
equations (\ref{intkin1}) if one sets $\dot{P}(\alpha;t)=0$ and $k_f=0$.
Since the relations $K_{0-}\simeq 0$ and $K_{0+}\simeq
0$ are valid  for the case of a charge transmission in the absence of an
applied voltage, it follows
$P(*;0)\simeq 0$, $P(+;0)\simeq 0$, $P(-;0)\simeq 0$ and $P(0;0)\simeq 1$.
As already indicated the hopping and the tunnel transition processes can
proceed in off-resonant or a  resonant regime depending on the sign of
the actual transmission gap,
cf. eqs. (\ref{gap+}) and (\ref{gap-}).

Although the calculations have been performed on the basis of the general
expressions (\ref{intkin1}), and (\ref{cseq2}) - (\ref{ineltun2}), including
eqs. (\ref{ineltun}), (\ref{phi}) and  (\ref{drate}) - (\ref{ineltun3}), most
of the findings will be discussed in terms of analytical expressions. Those
are derived for cases where an electron transfer occurs preliminary
along a separate transmission channel. For the sake of definiteness, we
consider a charge transmission process where the energy $E_--E_F$ is
larger than $E_*$ and $E_++E_F$ (see the cases (a) and (c) in Fig. \ref{fig06}).
We also suppose that $\Delta E_{-*}$ is large enough to neglect the population
of the state $M_-$. This means that the off-resonant and resonant regimes of
light-induced current formation involve an electron transfer process
predominantly across the three molecular states, $M_0$, $M_*$ and $M_+$
(the left route in Fig. \ref{fig05}).
Thus, charge transmission only occurs along the channel related to the
charged molecular state $M_+$. To achieve an analytic description of this
transmission process one has to set $K_{*-}\simeq 0$. This leads to the
following solution of the eqs. (\ref{intkin1}):
%
\begin{displaymath}
P(0;t)\simeq P_0 +\frac{k_f}{k_1k_2(k_1-k_2)}\,
\end{displaymath}
\begin{displaymath}
\times\big[k_2(k_1-\lambda_+-K_{*+})e^{-k_1t}
-k_1(k_2-\lambda_+-K_{*+})e^{-k_2t}\big]\,,
\end{displaymath}
\begin{displaymath}
P(*;t)\simeq P_* +\frac{k_f}{k_1k_2(k_1-k_2)}\,\big[-k_2(k_1-\lambda_+)\,e^{-k_1t}
\end{displaymath}
\begin{displaymath}
+k_1(k_2-\lambda_+)\,e^{-k_2t}\,,
\end{displaymath}
\begin{displaymath}
P(+;t)\simeq P_+ +\,\frac{k_fK_{*+}}{k_1k_2(k_1-k_2)}
\,\big[k_2\,e^{-k_1t}-
k_1\,e^{-k_{2}t}\big]\,,\;\;\;
\end{displaymath}
\begin{equation}
P(-;t)\simeq 0 \; .
\label{pop2g}
\end{equation}
The quantities
\begin{displaymath}
P_0=(K_{*+}K_{+0}+k_d\lambda_+)/k_1k_2
\end{displaymath}
\begin{displaymath}
P_+=k_fK_{*+}/k_1k_2\,,\;\;\;\;P_-=0\,,
\end{displaymath}
\begin{equation}
P_*=k_f\lambda_+/k_1k_2\,
\label{spop2}
\end{equation}
are steady state populations and the overall transfer rates take the form
%
\begin{equation}
k_{1,2}=(1/2)\big[\,a \pm\sqrt{a^2-4b^2}\,\big] \; .
\label{overa}
\end{equation}
Note the abbreviations
$a=k_f+ \lambda_+ +\lambda_*$, $b^2=k_f(\lambda_+ +K_{*+})+\lambda_{*}K_{+0}
+2k_dK_{+*}$, and
$\lambda_+ \equiv K_{+0}+2K_{+*}$, $\lambda_* \equiv K_{*+}+k_d\,$
Based on the eqs. (\ref{cseq2}) - (\ref{pop2g}) one can derive analytic expressions for the
sequential and direct current components. The time evolution of these components is
determined by the overall transfer rates $k_1$ and $k_2$.
Obviously, the formation of a finite photocurrent in the absence of an
applied voltage only becomes possible at an {\it asymmetric} coupling of
the molecule to the electrodes. As it was already noted such asymmetry can
result from  nonidentical electron density at the HOMO and the LUMO (cf.
Fig. \ref{fig01}). For the following we
assume $\Gamma^{(1)}_{H}> \Gamma^{(2)}_{H}, \Gamma^{(2)}_{L}>\Gamma^{(1)}_{L}$.
Since the factor $\Gamma^{(1)}_{H}\Gamma^{(2)}_{L}-
\Gamma^{(2)}_{H} \Gamma^{(1)}_{L}$
becomes positive the steady state electron current is also a positive
(electrons move from electrode 1 to electrode 2).

Next let us consider the current
formation at a weak molecule-electrode coupling (when the width parameters
are of the order $(10^{-7}-10^{-4})$ eV) and  at a moderate optical excitation
($k_f = 10^{10}$ s$^{-1}$). The excitation energy is $\hbar\omega=E_*-E_0=1.6$ eV.
The various parameters are collected in Table 1.
\begin{table}[h!]
\begin{center}
\begin{tabular}{||p{0.1\columnwidth}||p{0.11\columnwidth}|p{0.11\columnwidth}|p{0.1\columnwidth}|p{0.1\columnwidth}|p{0.1\columnwidth}|p{0.1\columnwidth}|p{0.1\columnwidth}||}
\multicolumn{8}{p{0.81\columnwidth}} {\bfseries Table I} \\
\multicolumn{8}{p{0.81\columnwidth}} {Parameters of the HOMO-LUMO
model ($\Delta E_{\pm *}$ and
$\Gamma^{(i)}_{j}$ are given in eV)} \\
\multicolumn{8}{c}{ } \\
 \hline \hline
 \centering Figs & \centering $\Delta E_{+*}$ & \centering $\Delta E_{-*}$ &
 \centering$\Gamma^{(1)}_{L}$ & \centering$\Gamma^{(2)}_{L}$ &
 \centering$\Gamma^{(1)}_{H}$ & \centering$\Gamma^{(2)}_{H}$
 & \multicolumn{1}{p{0.1\columnwidth}||}{\centering $T,^{o}K$}  \\
\hline \hline
 \centering 7 & \centering 0.4 & \centering 0.8 & \centering $10^{-6}$  & \centering$10^{-4}$ & \centering $10^{-5}$
 & \centering$10^{-6}$ & \multicolumn{1}{p{0.1\columnwidth}||} {\centering 300} \\
\hline
 \centering 8 & \centering 0.1 & \centering 0.8 & \centering $10^{-6}$  & \centering$10^{-4}$ & \centering $10^{-5}$
 & \centering$10^{-6}$ & \multicolumn{1}{p{0.1\columnwidth}||} {\centering 300} \\
\hline
 \centering 9 & \centering 0.1 & \centering 0.8 & \centering $10^{-6}$  & \centering$10^{-4}$ & \centering $10^{-5}$
 & \centering$10^{-6}$ & \multicolumn{1}{p{0.1\columnwidth}||}{\centering 100} \\
\hline
 \centering 10 & \centering -0.1 & \centering 0.8 & \centering $10^{-7}$ & \centering $10^{-5}$ & \centering $10^{-6}$
 & \centering $10^{-7}$ & \multicolumn{1}{p{0.1\columnwidth}||}{\centering 300} \\
\hline
 \centering 11 & \centering 0.8 & \centering -0.1 & \centering $10^{-7}$ & \centering $10^{-5}$ & \centering $10^{-6}$
 & \centering $10^{-7}$ & \multicolumn{1}{p{0.1\columnwidth}||}{\centering 300} \\
\hline
 \centering 12 & \centering -0.1 & \centering -0.2 & \centering $10^{-7}$ & \centering $10^{-5}$ & \centering $10^{-6}$
 & \centering $10^{-7}$ & \multicolumn{1}{p{0.1\columnwidth}||}{\centering 300} \\
\hline \hline
\end{tabular}
\end{center}
\end{table}
\subsection{Off-resonant regime of charge transmission}
Figs. \ref{fig07} - \ref{fig09} demonstrate the transient behavior of the
direct and sequential current components for a positive transmission gap
$\Delta E_{+*}$ (it determines the efficiency of the $M_*\rightarrow M_+$
transition due to thermal activation).

\subsubsection{Deep off-resonant regime}
\begin{figure}
\includegraphics[width=7cm]{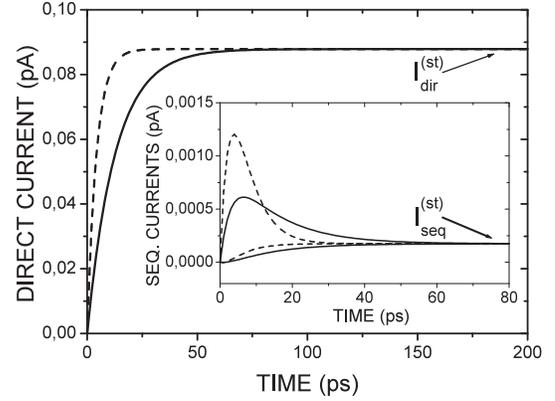}
\caption{Transient current components for an off-resonant charge transmission process
and at a different intensities of optical excitation. Solid lines: $k_f=10^{10}$ s$^{-1}$, dashed
lines: $k_f=3\cdot 10^{10}$ s$^{-1}$. Insert: sequential
current components $I_{seq}^{(1)}(t)$ (upper curves) and $I_{seq}^{(2)}(t)$
(lower curves) approach the common steady state value.
}
\label{fig07}
\end{figure}

In the case represented in Fig. \ref{fig07},  the gap $\Delta E_{+*}$ is
assumed to be a rather large so that
the population of the charged molecular state $M_+$ is small. As a result,
charge transmission mainly follows the tunneling route whereas thermal
activation of the molecular state $M_+$
(as well as state $M_-$) is suppressed (cases (a) and (b) of Fig. \ref{fig06}).
For such a transmission the distant current component strongly exceeds
the sequential one (see the insert). Therefore, the total current,
eq. (\ref{curdet1}) is associated with the direct component:
%
\begin{equation}
I_r(t)\simeq I_{dir}(t)\simeq I^{(st)}_{dir}
\,\big(1-e^{-t/\tau_{st}}\big)\,.
\label{off}
\end{equation}
Here, $\tau_{st}=k_{st}^{-1}=(k_f+k_d)^{-1}$ is the characteristic
time of the evolution of the current to its steady state value
%
\begin{equation}
I^{(st)}_{dir}=I_0\pi\hbar\,S_{*0}\,
(k_f/k_{st})\,
\label{curoffst}
\end{equation}
The quantities $S_{*0}$  and $Q_{*0}$  (the latter enters $k_d$, cf.
eq. (\ref{drate})) have been defined in the
eq. (\ref{dirtr+}) and eq. (\ref{ineltun2}), respectively, additionally using
$R \simeq (1/\pi)[(\Delta E_{+*})^{-1} + (\Delta E_{-*})^{-1}]$.

In the off-resonant regime, the inequality $k_f\gg Q_{*0}$ is valid with a
good accuracy. As a result,
a dependence of the photocurrent on the light intensity (i.e. on the rate
$k_f$) is only present within the transient behavior whereas the steady
state value, $I_{st}=I_{dir}^{(st)}$,
becomes independent on $k_f$ (see Fig. \ref{fig07}). We also note that the
single-exponential kinetics correctly describes the transfer
process until the population of the charged molecular state $M_+$ (and $M_-$)
becomes so small that the direct current component strongly exceeds the
sequential one.

\subsubsection{Single-channel off-resonant  regime}

\begin{figure}
\includegraphics[width=7cm]{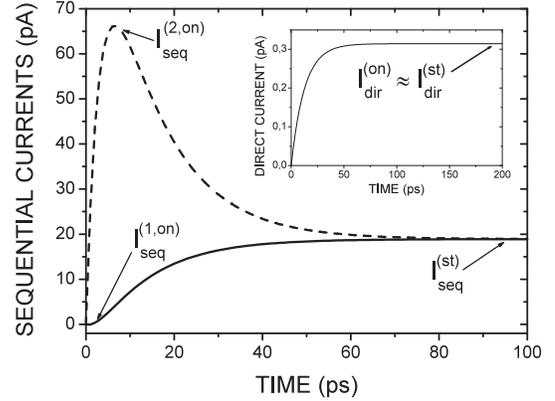}
\caption{Off-resonant regime of current formation at a small transmission gap $\Delta E_{+*}$.
}
\label{fig08}
\end{figure}

\begin{figure}
\includegraphics[width=7cm]{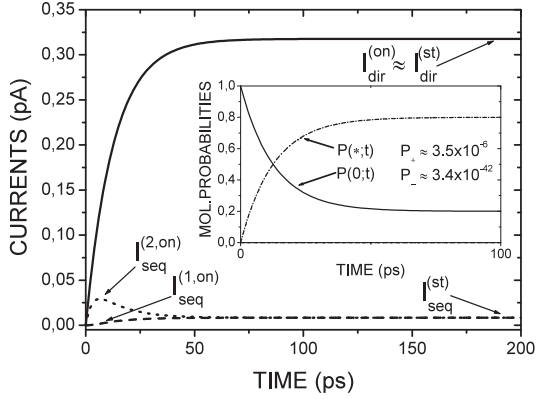}
\caption{Off-resonant current formation at low temperature with the
participation of the charged molecular state $M_+$.
}
\label{fig09}
\end{figure}

The Figs. \ref{fig08}-\ref{fig10} depict how the current components
approach their steady-state values
for the case of a large and positive gap $\Delta E_{-*}$ and a positive but not so large gap
$\Delta
E_{+*}$. Here, an activation of the hopping process $M_*\rightarrow M_+$ is
possible and the contribution of the sequential component to the total one becomes
significant.
As a result, the transient behavior of the current represents two-exponential
kinetics. For instance, a comparison of Fig. \ref{fig07} and Fig.
\ref{fig08} shows that at a small gap $\Delta E_{+*}$ (but at the same width
parameters and temperature), an electron transmission along the
sequential route becomes  more effective than the transmission along the
tunnel route. Therefore, the total current is not caused by the direct
component (as in Fig. \ref{fig07}) but by the sequential one (cf.
Fig. \ref{fig08}). As a result, the total currents
$I_1(t)\simeq I_{seq}^{(1)}(t)$ and
$I_2(t)\simeq I_{seq}^{(2)}(t)$ do not coincide in the transient region.
Such a behavior is originated by an asymmetric hopping of electrons
between the molecule and the electrodes.
A temperature decrease does not predominantly impact  the direct component
but strongly reduces the
sequential one, as deduced from a comparison of  Fig. \ref{fig09} and
Fig. \ref{fig08}.

For a further inspection of the transient current either in the off-resonant
or the resonant regime we use analytic expressions which follow from
the eqs. (\ref{cseq2}),  (\ref{cdir2}) and (\ref{pop2g}).
Let us start with an analysis of the sequential current components
%
\begin{displaymath}
I_{seq}^{(r)}(t)=I_{seq}^{(st)}\,\big[1-\frac{1}{k_1-k_2}\big(k_1e^{-k_2t}
-k_2e^{-k_1t}\big)\big]
\end{displaymath}
\begin{equation}
+(-1)^{r}\,I_0\pi\frac{k_f\Gamma_L^{(r)}}{k_1-k_2}N(\Delta E_{+*})\,\big(e^{-k_2t}
-e^{-k_1t}\big)\,
\label{cseq3}
\end{equation}
where the quantity
%
\begin{equation}
I_{seq}^{(st)}=I_0\pi\frac{k_f}{\hbar k_1k_2}(\Gamma_H^{(1)}\Gamma_L^{(2)}
-\Gamma_H^{(2)}\Gamma_L^{(1)})\,N(\Delta E_{+*})\,
\label{cseqst}
\end{equation}
denotes the steady state sequential current entering $I_{seq}^{(1)}(t)$ and
$I_{seq}^{(2)}(t)$.

The results depicted in the Figs. \ref{fig08} and \ref{fig09} refer to an
electrode-molecule coupling which guarantees $k_1\gg k_2$ where
$k_1\simeq k_f+k_d$ and $k_2\simeq  (1/\hbar)\,
[\Gamma_L\,(2-N(\Delta E_{+*})) +\Gamma_H]$.
Due to the inequality $k_1\gg k_2$ the fast and the slow kinetic periods of
the time evolution are well determined. This allows one to distinguish
between the currents $I_{seq}^{(1)}(t)$ and $I_{seq}^{(2)}(t)$.  The fast
kinetic phase covers  a time region of
the order of $k_1^{-1}$ and starts just after the switching on of the optical
excitation. The phase
ends at $t\gtrsim 5 k_1^{-1}$ from which the time-dependent behavior of
the current is determined by the slow phase
%
\begin{equation}
I_{seq}^{(r)}(t)\simeq I_{seq}^{(st)}\,\big(1-e^{-t/\tau_{st}}\big)+
I_{seq}^{(r,on)}\,e^{-t/\tau_{st}}\,.
\label{cseq4}
\end{equation}
Here,  $\tau_{st}=k_2^{-1}$ is the characteristic time the current needs to
achieve its  steady state value. The expression
%
\begin{equation}
I_{seq}^{(r,on)}= (-1)^rI_0\pi\Gamma^{(r)}_{L}N(\Delta E_{+*})\,
\label{ecur}
\end{equation}
gives the current component valid at $t\ll\tau_{st}$.
The sign of the $I_{seq}^{(r,on)}$ is determined by the direction of electron
motion from the photoexcited molecule to the $r$th electrode. Since in the
case under consideration, the transmission channel is associated with the
charged molecular state $M_+$, the $M_*\rightarrow M_+$ transition involves
an electron which leaves the LUMO and is captured by either
the 1st or the 2nd electrode. In the scheme depicted in Fig. \ref{fig05}
this transition is accompanied by an electron hopping from the LUMO to
electrode 1 and is characterized by the contact rate constant $K_{*+}^{(1)}$.

As a quantifier of the transient kinetics we introduce the ratio
%
\begin{equation}
\eta^{(r)}_{seq}=|I_{seq}^{(r,on)}|/I_{seq}^{(st)} \; .
\label{ratseq}
\end{equation}
It indicates how strongly the sequential components differ from their
steady state value if the optical excitation is switched on.
Our studies show that the difference between these quantities can become
large if the difference between the width parameters is large. We illustrate
this observation for the case $\Gamma^{(1)}_{H}\gg
\Gamma^{(2)}_{H}$ and $\Gamma^{(2)}_{L}\gg\Gamma^{(1)}_{L}$ so that $\Gamma^{(1)}_{H}
\Gamma^{(2)}_{L}-\Gamma^{(2)}_{H} \Gamma^{(1)}_{L}\approx\Gamma^{(1)}_{H}\Gamma^{(2)}_{L} $.
Therefore, if, for instance, $\Gamma^{(1)}_{H}=0.1\Gamma^{(2)}_{L}$ then
$\eta^{(r)}\simeq (\Gamma^{(r)}_{L}/\Gamma^{(1)}_{H})\big(2-N(\Delta E_{+*})\big)$ and thus
$|I_{seq}^{(1,on)}|\ll I_{seq}^{(st)}$
whereas $I_{seq}^{(2,on)} \gg I_{seq}^{(st)}$ , cf. Fig. \ref{fig08}.

During the slow kinetic phase the behavior of the direct current component is described by the
expression
%
\begin{equation}
I_{dir}(t)\simeq I_{dir}^{(st)}\,\big(1-e^{-t/\tau_{st}}\big)+
I_{dir}^{(on)}\,e^{-t/\tau_{st}}\; .
\label{cdir3}
\end{equation}
The two current components,
%
\begin{equation}
I_{dir}^{(st)}=I_0\pi\hbar\,S_{*0}\,\Big[1-\frac{\Gamma_L}{\hbar k_2}N(\Delta E_{+*})\Big] \;
\label{cdir3s}
\end{equation}
and
%
\begin{equation}
I_{dir}^{(on)}=I_0\pi\hbar\,S_{*0}\,,
\label{stdir}
\end{equation}
represent the steady state values, respectively. The basic difference in the
behavior of the direct and sequential current components is as follows: In
the off-resonant regime of electron transfer (where $N(\Delta E_{+*})\ll 1$)
the maximal value of the direct switch-on current coincides with its steady
state value, $I_{dir}^{(on)}\simeq I_{dir}^{(st)}$.
Therefore, a slow kinetic phase is not observed for the direct  component.
The respective time evolution is determined by the fast
single exponential kinetics (see Figs. \ref{fig07} and \ref{fig09}).  This behavior is related to
the small population of the charged molecular states. The conclusion is that
at an off-resonant regime, the appearance of a large switch-on current (in comparison to its
steady state value) could be only related to one of the sequential components
but not to the direct component (compare Figs. \ref{fig07} and \ref{fig09} with Fig. \ref{fig08}).

\subsection{Resonant regime of charge transmission}

The resonant regime of charge transmission is formed if the transmission
gaps $\Delta E_{+*}$ and $\Delta E_{-*}$ become negative. The temporal evolution
of the respective current components is represented in the Figs.
\ref{fig10} - \ref{fig12}. The used light intensity is identical with the one taken for the study of
the off-resonant regime.

\subsubsection{Single-channel resonant regime}
\begin{figure}
\includegraphics[width=7cm]{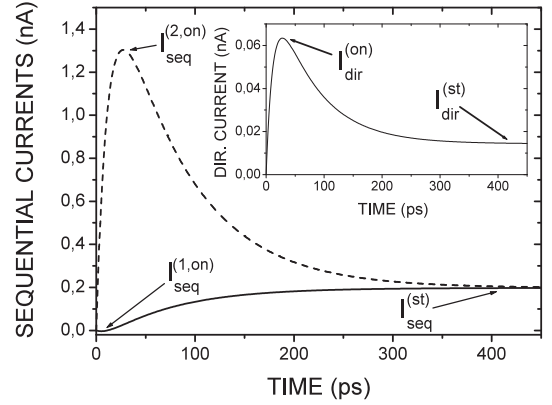}
\caption{Single-channel resonant charge transmission with the participation of the charged
molecular state $M_+$ (the energy gap $\Delta E_{+*}$ is negative).
The total current is mainly determined by the sequential components.
}
\label{fig10}
\end{figure}

\begin{figure}
\includegraphics[width=7cm]{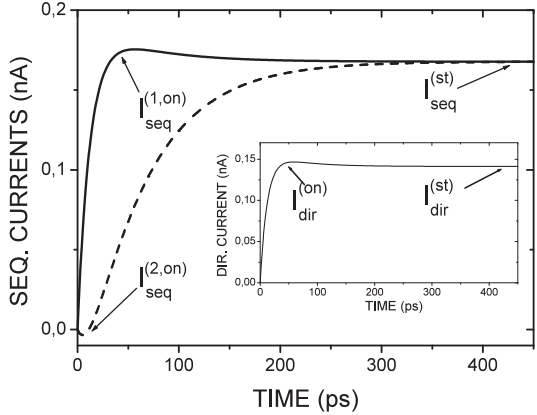}
\caption{Single-channel resonant transmission with the participation of the
charged molecular state $M_-$.
}
\label{fig11}
\end{figure}

This regime is achieved if $E_*>E_+ +E_F$ or $E_*+E_F>E_-$, i.e. if
$\Delta E_{+*}<0$ or $\Delta E_{-*}<0$, respectively (cases (c) or (d) of
Fig. \ref{fig06}). Physically, the cases $\Delta E_{+*}<0$, $\Delta E_{-*}>0$
and $\Delta E_{+*}>0$, $\Delta E_{-*}<0$ do not differ from each other.
Therefore, for the sake of definiteness, let us analyze the case (c). As in
the previous subsection, a situation is considered where the fast and
the slow kinetic phases are clearly differ from each other. Accordingly, the
relaxation of the current components to their steady state values is
described by eqs. (\ref{cseq4})-(\ref{stdir}) where now
$\Delta E_{+*}<0$ and, thus,
%
\begin{displaymath}
S_{*0}=\frac{1}{\hbar\Gamma_L}\,\big(\Gamma^{(1)}_{H}
\Gamma^{(2)}_{L}-\Gamma^{(2)}_{H}\Gamma^{(1)}_{L}\big)\,,
\end{displaymath}
\begin{equation}
Q_{*0}=\frac{1}{\hbar\Gamma_L}\,\big(\Gamma^{(1)}_{H}
\Gamma^{(2)}_{L}+\Gamma^{(2)}_{H}
\Gamma^{(1)}_{L}\big)\,.
\label{offst1}
\end{equation}
A comparison of Fig. \ref{fig10} and Fig. \ref{fig08} shows that the change of $\Delta E_{+*}$
from positive values ($\Delta E_{+*}=0.1$ eV) to negative ones ($\Delta E_{+*}=-0.1$ eV)
results in a significant increase of the sequential and distant current components (despite the fact
that  the width parameters are taken even less than those used in the Figs \ref{fig08} -
\ref{fig10}). At the same time, the ratio (\ref{ratseq}) between the switch-on and the
steady state sequential current component is conserved. In the case of
the resonant regime, one can also introduce the ratio
defined by the direct current component:
%
\begin{equation}
\eta_{dir}=I_{dir}^{(on)}/I_{dir}^{(st)}\,.
\label{ratdir}
\end{equation}
In line with the expressions (\ref{cdir3s}) and (\ref{stdir}) it yields
$\eta_{dir}=\Gamma_L/\Gamma_H$.
If one takes the same relation between the width parameters as it has been
used in Fig. \ref{fig08}, then
$\eta_{dir}\simeq \Gamma_L^{(2)}/\Gamma_H^{(1)}\sim 10$
in correspondence with the exact results depicted in Fig. \ref{fig10}.
The above given results refer to a current formation connected with the
transmission along the channel which is associated with the charged molecular
state $M_+$. The channel includes two types of transmission routs depicted
in Fig. \ref{fig06}, the left sequential route ($M_*\rightarrow
M_+\rightarrow M_0$) and the direct tunnel route ($M_*\rightarrow M_0$).
Analogously, one can consider the current formation if a charge transmission
occurs preliminary along the channel related to the molecular charged state
$M_-$ (right sequential  route $M_*\rightarrow M_-\rightarrow
M_0$ and direct tunnel route $M_*\rightarrow M_0$). Recall that the distant
rate constants $Q^{(12)}_{*0}$ as well as the $Q^{(21)}_{*0}$ are defined by
both charged molecular states $M_+$ and $M_-$.  To derive respective analytic
expressions, one sets
$K_{+*}=0$ in the eqs. (\ref{cseq2}), (\ref{cdir2}), and  (\ref{contrate}).
This results in an analytic form which follows from eqs. (\ref{pop2g})  and
(\ref{stdir}) if one replaces $K_{-*(0)}$ and $K_{*(0)-}$ by $K_{+*(0)}$ and
$K_{*(0)+}$, respectively. Our studies show that in this case,  the
direct current component is comparable with the sequential components
(see  Fig. \ref{fig11}). Since the fast regime of charge transmission is associated now
with the hopping of an electron into the HOMO,  the  maximal value of the sequential current
component is less than that of Fig. \ref{fig10}due to the condition  $\Gamma_H^{(1)} <
\Gamma_L^{(2)}$.

\subsubsection{Two-channel resonant regime}

\begin{figure}
\includegraphics[width=7cm]{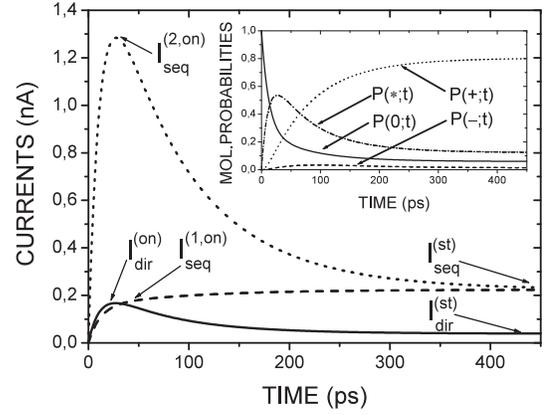}
\caption{Two-channel resonant charge transmission. There is no basic difference between the
behavior of the current components formed by a single-channel of charge transmission
(except a certain increase of the direct component, cf. also Fig. \protect\ref{fig10}). The
evolution of the current components follow the time-dependent behavior of the molecular
probabilities (see insert).
}
\label{fig12}
\end{figure}

This regime is realized if both transmission channels
associated with the molecular charged states $M_+$ and $M_-$ participate in
the electron transfer process and if the respective transmission gaps
$\Delta E_{-*}$ and $\Delta E_{+*}$ are negative (cases (d) and (e) in
Fig. \ref{fig06}). In this two-channel resonant regime, the left and the
right electron transfer channels represented in Fig. \ref{fig05}, give a
comparable contribution to the current. The time-dependent evolution of the
current components is described  by the general expressions (\ref{cseq2})
and (\ref{cdir2}) and the set of kinetic equations (\ref{intkin1}).
Fig. \ref{fig12} does not show any different behavior among
the particular currents which belong to a particular channel.
The insert of Fig. \ref{fig12} demonstrates that the fast  part of the time evolution
completely
corresponds to the  kinetics of formation of the excited molecular state
$M_*$. The slow part reflects the kinetics at which a population of the
charged molecular states $M_+$ and $M_-$ varies due to a depopulation
of the light-induced state $M_*$. Such a depopulation is negligible during an
off-resonant regime of charge transmission but  becomes pronounced  in
the resonance regime.

\section{Conclusions}

We put forward a detailed  study  on the time-dependent   behavior of
light-induced transient currents in molecular junctions, like in a molecular
diode \textbf{1}-M-\textbf{2}. A nonequilibrium  set of kinetic equations
has been derived for the molecular states which participate in the current
formation. Those are the neutral molecular states $M_{0}$ and $M_{*}$ and
the two charged molecular states $M_{+}$ and $M_{-}$. It could be shown that
an interelectrode electron transfer
$\textbf{1}\textbf{2}\rightleftarrows \textbf{1}^+\textbf{2}^-$ takes place
along the channels associated with the charged molecular states $M_+$  and
$M_-$,  see Fig. {\ref{fig05}. These states participate in a light-induced
interelectrode electron transfer in the absence of an applied voltage either
as  real intermediate states (forming the sequential transmission route) or
as  virtual intermediate states (forming the direct  transmission route).
The sequential route includes the hopping of an electron between the molecule
and the adjacent electrodes, being thus responsible for a molecular charging.
The formation of the respective direct current component is accompanied by
the transition of the molecule from its photoexcited state $M_*$
to its ground-state $M_0$. The time-dependent  behavior of the total
photocurrent is governed by the molecular  populations $P(\alpha;t)$,
cf. eqs. (\ref{cseq2})  and (\ref{cdir2}). The latter evolve
in line with the kinetic rate equations (\ref{intkin1}).

The relative efficiency of each route is determined by the actual value
of the transmission gaps $\Delta E_{\alpha*}$,  ($\alpha =+,-$): If
$\Delta E_{\alpha*}$ is positive, then an electron transmission along
the $M_{\alpha}$-channel proceeds in an off-resonant regime whereas for a
negative  $\Delta E_{\alpha*}$ the current is formed in the resonant regime.
In this regime the sequential and the distant current components significantly
exceed the same components formed at the off-resonant charge transmission
(see in this context the Figs. \ref{fig10} and \ref{fig11}
and compare them with the Figs. \ref{fig07} and \ref{fig08}). In the framework of a
HOMO-LUMO model, the direction of the light-induced electron
current is determined by the factor $\Gamma^{(1)}_{H}
\Gamma^{(2)}_{L}-\Gamma^{(2)}_{H}
\Gamma^{(1)}_{L}$ which reflects the  difference between interelectrode
electron flow $\textbf{1}\textbf{2}\rightarrow \textbf{1}^+\textbf{2}^-$
and $\textbf{1}^-\textbf{2}^+\leftarrow \textbf{12}$.

If the difference between the width parameters $\Gamma^{(r)}_{j}$ becomes
large, a characteristic kinetic effect appears for the transient photo
currents. In this case, those currents do exceed their steady state value to
a large amount. The physical origin of this effect is related to the fact
that the photoexcitation of the molecule ($M_0\rightarrow M_*$ process)
initiates a molecular charging (see
the schemes in Fig. \ref{fig04} and Fig. \ref{fig05}). Charging is caused by the transition
of an electron from the molecule to each electrode ($M_*\rightarrow M_+$ process)
or from the electrodes to the molecule ($M_* \rightarrow M_-$ process).
Such a light-induced electron motion
forms the fast initial part of the electron transmission which can be seen in the  time-dependent
behavior of the sequential current components $I^{(1)}_{seq}(t)$ and $I^{(2)}_{seq}(t)$.
Molecular recharging is characterized by contact (hopping) rate constants. If the characteristic
recharging
time is  much less than the characteristic time $\tau_{st}$ of the steady state formation, then the
maximal value of the transient photocurrent may become rather large compared to its steady
state value (see, for instance, Fig. \ref{fig10} and Fig. \ref{fig12}).

Experimental studies of  transient photocurrents allow one to clarify the details of contact
(molecule-electrode) and distant (electrode-electrode) electron transfer processes  in  molecular
junctions. We have  shown that the effective formation of the photocurrent becomes possible if
the photon energy $\hbar\omega =E_*-E_0$ exceeds the energy  gaps $\Delta
E_{-0}$ or/and $\Delta E_{+0}$, cf. Fig. \ref{fig02}). This corresponds to an energy level
arrangement as shown in the schemes (c)-(f) of Fig. \ref{fig05}.

\section {Acknowledgments}
E.G.P. gratefully acknowledge generous support by the Alexander von Humboldt-Foundation.
P.H. and V. M. acknowledge the support by the Deutsche Forschungsgemeinschaft (DFG)
through priority program DFG-1243 {\it Quantum transport at the molecular scale} (P. H.) and
through the collaborative research center (Sfb) 951 {\it Hybrid inorganic/organic systems for
opto-electronics} (V. M.).




\begin{thebibliography}{00}
\bibitem{foot1} The wide-band approximation \protect\cite{nit01} is valid for electrodes fabricated from the noble metals. In this case, the width parameter (\protect\ref{width1}) can be choosen as not exhibiting a dependence on the transmission energy $E$.
\bibitem{am} While the current unit is given in Amperes one has to set
    $\hbar\simeq 6.58\cdot 10^{-16}$eV$\cdot$s. This means that the quantities $\hbar K_{\alpha\alpha'}$, $\hbar Q^{(rr')}_{*0}$, and the width parameters $\Gamma_j^{(r)}$ are taken in eV.
\end{thebibliography}
\end{document}